# Secure Non-public Health Enterprise Networks

Mona Ghassemian
*Applied Research*
*British Telecommunications plc*
*Adastral Park,*
*Ipswich, IP5 3RE, UK*
mona.ghassemian@bt.com

Max Smith-Creasey
*Applied Research*
*British Telecommunications plc*
*Adastral Park,*
*Ipswich, IP5 3RE, UK*
max.smith-creasey@bt.com

Maziar Nekovee
*Engineering and Design*
*University of Sussex*
*Sussex House, Falmer*
*Brighton, BN1 9RH, UK*
m.nekovee@sussex.ac.uk

*Abstract*—Increasing demand for secure remote operation in industry and technology advancements to support delivering efficient services and tele-mentoring have opened a new market in healthcare sector and emergency services based on 5G and Tactile internet capabilities. In a connected world, hospitals would benefit from providing the on-time availability either for continuous health monitoring or critical services to the citizens in need. In this paper, we propose a secure non-public health enterprise network concept to enable an end-to-end secure and location-agnostic communication between a patient and a healthcare service provider, and other contacts with patient's consent either in case of an emergency or to be stored in the medical records. We present how applying non-public enterprise networks can address market demand in health care sector for improved end-to-end security and privacy when dealing with personal and critical information. We present the three-tier architecture model describing continuous authentication mechanisms based on biometric collection as well as the dynamic network solutions in the healthcare domain. The biometric collection can be done using ambient/IoT sensors as well as wearable/implantable devices to monitor the patient unobtrusively. Furthermore, end-to-end security solutions should adapt dynamically based on the user profile and situation awareness to address the required level of security at the network side. We discuss the related research challenges for developing the presented non-public health enterprise platform and provide suggestions for future work based on the healthcare sector requirements and opportunities.

*Keywords—dynamic security; health IoT; continuous authentication; biometrics; non-public enterprise networks; location agnostic; 5G*

## I.    INTRODUCTION

In recent years there have been significant advancements within the Internet of Things (IoT) field. These advances have beneficial application to healthcare in order to monitor and communicate via a variety of sensors and devices. Through IoT monitoring devices patients can live with greater regularity without the inconvenient and restrictive task of explicitly monitoring their health conditions. The use of IoT further lightens the workload of medical professionals by providing them with immediate and accurate data from which decisions can be made without patient interaction.

The traditional healthcare system follows a rigid and inflexible architectures in which patients report their activities and health to a medical professional through direct contact. Through greater implementation of IoT devices in the healthcare domain there can be greater benefits to patients and to medical professionals. In this paper, we describe and illustrate these benefits of IoT assisted healthcare. Though, as will all connected devices, especially within the healthcare field, security and privacy are of paramount importance to ensure the data is not obtained by third parties and remains confidential between a patient and their medical professional. To this end, a non-public enterprise network is discussed for the communication of this data within the healthcare domain. The devices used by medical professionals to access the healthcare data of their patients must also be secure. The security of these devices, however, is often managed by crude one-shot passwords which are inconvenient as well as insecure. In January 2020, it was found by BBC News that staff in the NHS must remember up to 15 different computer logins [1]. We apply state-of-the-art continuous authentication concepts as a solution to provide a greater level of convenience, usability and security.

Beside the access level, security solutions are required to insure a secure and reliable communication to all authenticated users. In January 2019, first remote operation using 5G was performed on an animal in China [3]. While such delay sensitive applications can be supported by ultra-reliability low latency (URLLC) capability in 5G networks and beyond, further security and reliability requirements derive such application to be run on non-public networks. In this paper, we present a non-public health enterprise network platform with a focus on the security and privacy of patient information. We define an end-to-end (e2e) dynamic security concept utilising IoT healthcare and discuss how different stakeholders and users of continuous health monitoring can benefit from the presented secure platform. We also describe how this data can be securely accessed at the medical professional end through continuous authentication techniques.

The rest of this paper is organised as follows: We present the related work to health IoT solutions, non-public networks and authentication solutions for health sector in Section II. Section III describes the market drive and requirements as well as a detailed description of the use cases considered in this work. We present the end-to-end dynamic security concept in Section IV, which includes both user continuous authentications as well as the networking solutions. Section V highlights the relevant research challenges. Finally, Section VI concludes the paper.

## II.    RELATED WORK

In this section we describe the studies related to different aspects of the proposed concept; healthcare use cases, continuous authentication, and 5G security solutions and non-public networks.

### A.    Healthcare

To cover all scenarios for a continuous health monitoring for medical record and intervention, in this subsection we present related works for use cases at home, hospital, on the road scenarios as illustrated in Figure 1.

Aftercare monitoring of patients after being discharged from hospital and observing their convalescence for recommended period by their physicians in their own space prevents either missing a critical situation that requires sending a premed at home for assistance or an unnecessary re-admission. Continuous health and activity monitoring systems can be devised by either ambient sensors located in home [14], hospital or ambulance or wearables/implantable carried by the patients as they move [15] on the roads. A number of smart environment projects with physical testbeds and trials have been implemented [18] and the resulting datasets are available for researchers to mine, including the CASAS project [19], Technology Integrated Health Management (TIHM) [21], or smart hospital process mining for process management [24].

Beside the monitoring systems, tele-surgery is another health use case in hospitals. Robotic surgery systems (such as intuitive da vinci and Cambridge Medical robotics) have set up the platform for a local surgeon in the operation room to operate using visual and haptic cues and get assistance from an expert colleague if required. Communication technology advancements such as 5G URLLC and Tactile Internet [20] (IEEE 1918.1) aiming to provide a closed loop control feature with latencies in the range of milliseconds have enabled tele-surgery over a certain remote distance.

### B. 5G and non-public networks

5G networks are designed to address the needs of the enterprise for use cases benefiting from network slicing and (URLLC) and edge computing, particularly for industrial automation, healthcare, and public services. A number of 5G commercial deployments are already in place around the world, providing 5G services to customers. In addition, a number of non-public 5G networks are being deployed, for example NPNs designed to be operated as 5G testbeds, e.g., 5G-VINNI [10]. 5G security solutions such as Security Edge Protection Proxy (SEPP) node which is an entity to terminate signalling messages between PLMNs [11], as well as network slicing and slice isolation improve 5G services e2e security. Further ongoing 3GPP standardisation developments (SA3) on URLLC, addresses the reliability aspects indicating that terminal sets up two redundant PDU sessions over disjoint user plane paths.

Non-public networks with a variety of deployments options provides further ssecurity and privacy to the enterprise [12] mainly due to the dedicated resources and flexibility for access control definitions. are fast becoming a reality as a result of the rising demands of the enterprise, the increased willingness of telecom regulators to set aside licensed spectrum for non-public networks, and the underlying enabling technologies. Furthermore, due to security and privacy reasons many enterprises, and in particular the health sector prefers their data to remain local and under their control rather than being routed to a public cellular network. Also, the requirements may differ in terms of time or location, highlighting the need for custom networks (i.e., non-public network solutions) to plan and be reconfigurable and responsive to the underlying needs of the enterprise.

### C. Continuous Authentication

Continuous authentication has gained considerable research traction in recent years due to the proliferation of devices that enable the collection of multi-modal biometrics via a variety of sensors (such as an accelerometer for gait analysis). However, the use of continuous authentication to benefit healthcare communication over non-public networks has not been sufficiently explored. This section describes the works relating to devices commonly used in healthcare.

Mobile devices are commonly used among healthcare professionals. The screens on mobile devices have shown to be rich sources of biometrics in a variety of studies. In [2] the authors present a novel scheme for touch-gesture based authentication on mobile devices, achieving an equal error rate (EER) of 0-4%. Typing with keystrokes [6] or gestures [4] has also been shown to yield promising biometrics for continuous authentication. The inclusion of cameras on most modern mobile devices has led to the realisation of continuous face authentication [17]. More recently, schemes employing multi-modal behaviours have been produced to make use of the user environment (e.g., wi-fi and Bluetooth information) to better authenticate [5]. Some schemes have employed similar biometrics to establish which access policy to permit the user [7]; something that can be used to provide a tiered access to data based on the privacy of that data. Laptops and desktop computers are also regularly used in the healthcare profession for the input and access of information. A popular keystroke study was introduced in [8] with EERs of ~10%. In [9], the authors show mouse movement behaviour is a feasible biometric to authenticate. As with mobile devices, the face has also been suggested for continuous authentication schemes on computers [16]. This section reveals that whilst there have been efforts made to continuously authenticate, such schemes have not seen significant application in healthcare.

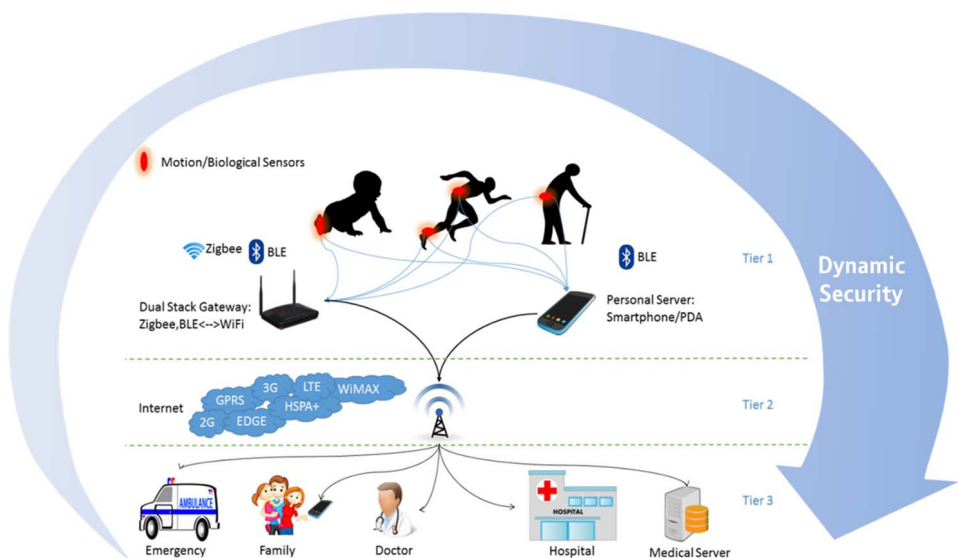

Figure 1: A three-tier health use case example

Figure 1. show a summary of the use case in which motion/biological sensors on patients (tier 1) relay information over the internet (tier 2) to the medical services (tier 3). Conversely, the authenticated medical devices (tier 3) can access the medical data of motion/biological sensors (tier 1) via the internet (tier 2).

## III. MARKET REQUIREMENTS AND USE CASE EXAMPLES

The generally accepted view of healthcare service providers is that they are slow adaptors of IoT health technology. Gartner's IoT survey [13] finds quite the opposite. The results show that the healthcare provider industry's IoT adoption is on par with retail and manufacturing industries. Health service provider decision makers have firmly embedded IoT into their operations, for both clinical care delivery and business processes.

### A. Market requirements

A public network is not designed for sensitive data and customised services that health sector services demand. Non-public networks (NPNs) provide many promising opportunities to communication service providers to deliver customised services. Strong demands for healthcare and public services and the personal data rights, patient health confidentiality, integrity and privacy demand IoT health solutions to comply with current security and privacy policies and procedures, namely GDPR and HIPPAA and breach rules updated by updated in 2013 by the HITECH Act. This impacts on patient experience in the reliability of the service and improves patient outcomes.

Beside the personal data sensitivity, reliable secure networks for use cases such as tele-surgery has been discussed for some time. There is an increase in the uptake of surgeons to move from laparoscopic to robotic surgery. Providing reliable and secure access to surgery in remote locations may require local surgeons to be assisted and mentored by experts remotely. This increases patient access to expert surgical interventions which will no longer be bound by geographical restrictions.

Another market driver need is the steady growth in the population of elderly. The number of elderly is expected to double by 2050, when it is projected to reach nearly 2.1 billion [18] worldwide, which results in more demand for care services for this generation. Ambient Assisted Living (AAL) mainly using IoT solutions address three major challenges; the ever-increasing global aging population, high expenses of geriatric health for government and the growing desire of elderlies to remain independent in their own home and aim to improve the Quality of Life (QoL) and well-being of elderly people to help them organise and manage their lifestyle more independently and within their comfort zone i.e., their homes.

### B. Use case examples

We categorise the healthcare use cases in three scenarios where the patient/other stakeholders are in medical premises, at their home, or outdoor between home and hospital:

- *Home:* Unobtrusive data collection in a smart home environment can monitor and assess residents' health and wellbeing based on ambient sensors or wearable devices carried by them. Patient wearable devices that send biometric data (such as blood pressure, heart, glucose level, and activity level monitoring devices) via a home monitoring device can assist with remote health monitoring. Ambient sensors embedded into the smart home (such as motion sensors and smart meters) as well as vision based monitoring solutions can capture readings from residents' daily routines to gain insights on human daily behaviour (such as activities, movements, gestures and identities) to evaluate the functional ability of residents for independent living. A common application scenario is the aftercare monitoring of an elderly to observe their recovery in their own space, immediately after being discharged from hospital for a duration suggested by their physicians.

- *Hospital:* Two categories of services can benefit from health IoT technology in hospitals and clinics with respect to their delay requirements:

  i) Delay sensitive applications: Proliferation of robotic surgery coupled with patients preference for minimally invasive/ reliable robotic surgery increased the spotlight on the lack of access to experienced surgeons and need for mentoring for novice robotic surgeons remotely. Currently surgeons operate with robots in one room (close vicinity) either using visual or simulation of haptic but the trust on the system reliability is an issue for remote tele-surgery. Tactile internet [20] can provide a "tele-mentoring" platform for robotic surgeons to improve patient outcomes and give all patients the possibility of access to the experts.

  ii) Delay tolerant applications: In addition to tele-surgery, IoT devices can be used to capture patients' record as well as improve hospital management efficiency. For instance, tracking the hospital assets using RFID sensors in a hospital can assist staff locating them when required. Furthermore, data collected by ambient devices tracking nurses' mobility pattern can be used by hospital management process to improve the work efficiency and resource management.

- *On the road:* Either when the patients is continuously being monitored or is transferred by emergence services to hospital, a location agnostic communication solution can connect a doctor in a hospital to utilise real-time information transmitted by a paramedic in the connected ambulance or at the incident site to help remotely guide/mentor and give necessary instructions and collect further information required (e.g., by voice/video and ultrasound images) to make decisions for the patient/casualty in acute care situation. Additionally, the information exchanged can help directing the ambulance driver to a local hospital with the necessary care units and available resources for the patient condition. The sensors that relay the telematics signals communicating the health status can be remotely visualised and controlled by the doctor providing a realisation of a tele-mentoring scenario. Beside emergency services, patients on the road can be monitored securely and continuously which gives them the freedom and assurance to live independently and run their day-to-day routine knowing if there is a need to call for assistance, their health record detects the anomaly and alert the hospital.

## IV. DYNAMIC SECURITY

In this section, we describe our presented three-tier secure non-public health enterprise nnetwork concept. We describe components, technologies and protocols across the tiers described in Figure 1 and demonstrates the interactions between them in Figure 2.

### A. Tier 1 – IoT in Healthcare

The first tier of our proposed secure non-public health enterprise network concept consists of IoT devices with medical capabilities that are prescribed to patients as part of their care package. As discussed, these devices collect information about the wellbeing of the patients in non-medical environments (e.g.: their home) and relay this information over technologies described in Tier 2 for medical professionals to access from devices at Tier 3. In IoT-assisted healthcare environments a variety of connected sensors monitor patients using smartphone applications communicating with wearables through Near Field Communication (NFC), such as the FreeStyle Libre blood glucose monitoring system, or over Bluetooth, such as Fitbit and Apple Watch with activity and heart-rate related healthcare capabilities. Home monitoring sensors such as Passive Infrared (PIR) sensors can be connected in a smart-home environment and pass information regarding user activity and movement to Tier 3 via Tier 2 through a home router. Wearable biosensors can also collect the biometrics which are translated to the physical and emotional state either in the local device or in a server connected to the system.

Due to the sensitivity of the personal and health data collected from mobile apps, wearables or implant sensors require secure communication links, data encryption and authentication techniques are required.

### B. Tier 2 – Non-public Networks

The dynamic edge-to-edge solution should encompasses the needs to have an appropriate level of security for the flow of information from home, or ambulance to the doctor in hospital in order to determine and monitor the patient conditions, in addition, sensors, devices and tools at home or on the road should be activated, controlled and operated both locally and remotely. This use case could also be extended by the ability (in the ambulance or home hub) to have an edge node capable of providing processing, storage, sensing/actuation, and communications capabilities. The node could additionally be extended by introducing voice-controlled commands. Non-public (private) Health Enterprise Network concept (Figure 2) is demonstrated with a focus on e2e dynamic/on-demand security for healthcare and emergency services relying on IoT platforms for health data collection from patient and people under continuous monitoring.

The proposed concept of secure non-public health enterprise network is based on non-public networks discussed in Section II.B. The customised/isolated nature of non-public networks specifically enables examining security requirements and providing security by design. The idea of customised cellular using the network slicing solutions and therefore the network slice isolation is in favour of e2e security. Additionally, in non-public networks no legal interception is required; hence no need to open the network, which may result in security concerns including:

- Privacy: Dedicated/customised non-public network deployment options allow users to have more control about their data storage and access level provision to ensure the data and network level security and sustain a competitive ecosystem of European technology and system providers in IoT as well as ensuring end-user trust, adequate security and privacy especially for the health use case where personal, physical and mental health data need to be collected from patients and individuals under monitoring.

- Mobility: Beside the health data, other type of communications can be relayed over the same transport system, interoperability between private and public deployment options is required to address security implications (e.g., ID, Authentication). Particularly when a user moves between a public to private network or if a device needs to connect to both simultaneously.

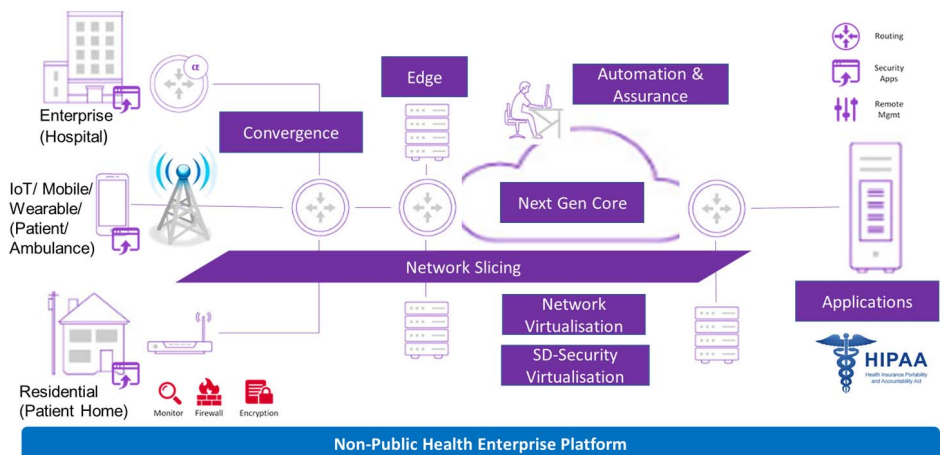

Figure 2: Non-public Health Enterprise concept for end-to-end dynamic security

- E2E security architecture: Furthermore, depending on the non-public enterprise network deployment options, the responsibility of security provisioning differs from the conventional service provision differs in non-public network architecture.

Network operators are also facing one very different kind of challenge to network security: Cyber criminals have made the cloud a tool for disrupting essential infrastructure services and compromising vulnerable assets. As a result, operators need faster, more efficient ways to identify and mitigate this distributed denial of service (DDoS) attacks. Real-time visibility and control can play a key role, by helping to disable threats and reroute traffic quickly. Dynamic security concept allows providing different type of security to the same user/device based on their type of service and adapt to human-centred IoT evolution improving usability and user acceptance, notably through strengthened security and user control. Dynamic security covers preventive security when in the design phase, but also can react to security attacks based on intent-based network segmentation to automatically identify the hardware and the services it is operating. Through continuous monitoring of the network, it is possible to identify sudden malicious behaviour which may compromise the performance of other services. These compromised units will be automatically isolated from their peers and they will continue to operate in isolation − which may lead to them providing only a reduced functionality, but they no longer can negatively impact the rest − until other MNO services resolve the issue and they automatically resume to their normal network service. The methodology for dynamic security relies on intent-based networking (learning from biology & physical security where by continuously monitoring (AI/ML learning what's normal), the SDN/NFV functional block of the 5G network is programmed to dynamically change policies, encryption, reroute, etc according to the use case requirement on the fly.

## C. Tier 3 – Continuous Authentication

In the third tier of the architecture there are all devices managed by medical professionals at hospitals, in ambulances and on their person. The use of continuous biometric authentication within the healthcare environment can greatly improve the experience for professionals and improve their utility. This is because such professionals would no longer need to spend time logging into a plethora of different systems but be automatically authenticated based on their biometrics. In order to comply with the market requirements of privacy and confidentiality for patients and their data it is crucial that strict state-of-the-art authentication and security mechanisms are put into place. Furthermore, it is important that the security is not inconvenient to the extent that it impedes the ability to offer adequate healthcare. The IoT devices often used by medical staff includes a variety of IoT devices but most commonly phones, laptops and tablets. Each of these devices contains a variety of sensors from which biometrics can be extracted (e.g.: smartphones often contain cameras, touchscreens, accelerometers, microphones, etc.). Our architecture therefore makes use of continuous authentication to authenticate the medical professionals based on continuous and passively collected biometrics. In all cases, the biometrics are collected locally, encrypted and sent to a service for decryption, feature extraction and authentication. The processing occurs on server (as opposed to on-device) such that the user can be logged for audit purposes, to prevent unnecessary battery drain and because it is expected that medical IoT devices will naturally have a persistent internet connection. Each IoT device can capture multiple biometrics. Once these biometrics are received by the authentication server, they will be passed to the relevant machine learning/statistical model (e.g.: convolutional neural network for face-based authentication). The subsequent score output by the models can then be fused via a biometric fusion technique [22], such as sum score fusion, defined as follows:

$$scr_{fused} = \frac{src_1 + src_2 + \cdots + src_n}{n}$$

This score $scr_{fused}$ can then influence a global trust score $T_G$. A reward, $r$ or a penalty, $p$ may be added dependant on if the score is greater, less than or equal to a threshold $thr$. The more extreme values for rewards and penalties, the more reactive the scheme. This adapted from [23] and defined as:

$$T_{G_{new}} = \begin{cases} \max(0, T_{G_{prev}} - p) & \text{if } scr_{fused} < thr \\ \min(1, T_{G_{prev}} + r) & \text{if } scr_{fused} > thr \\ T_{G_{prev}} & \text{if } scr_{fused} = thr \end{cases}$$

The new trust score [23] for the medical IoT device is then compared to thresholds that indicate whether the defined policy allows for continued access to the device. The decision can then be relayed from the server back to the device. As discussed, with maintained authentication the device then enables access over Tier 2 to the patient devices in Tier 1 as shown in Figure 1.

## V. RESEARCH CHALLENGES

This section discusses the research challenges that would enhance the proposed architecture within modern healthcare.

### A. IoT in Healthcare

Security and trust as well as communication latency are key challenges for delivering tele-surgery. Currently, conventional internet security protocols (namely, AES, WEP, WPA) are used which makes the data transfer prone to attack. Security and reliability of the data (including haptic) locally/remotely are key to telesurgery use-cases. Further work is required on security/privacy aware haptic data/feedback encoding techniques to improve the reliability and security of the tele-surgery. Furthermore, continuous monitoring demands low-power and reliable operation to avoid any interruption in data collection from power restricted medical devices and therefore the service delivery.

### B. Non-public Networks

Considering the sensitive health information and demand for high reliability in health operation, communication security and data privacy demand from healthcare service providers can be address by using non-public enterprise networks. The current standard and deployment options [25] however do not consider further security and reliably required by health-IoT and remote tele-surgery scenarios.

While the idea of customised cellular using the network slicing solutions and the network slice isolation is in favour of e2e security, when access to multiple slices are required the exposure of the slices need to avoid security and privacy breeches. The customised/isolated nature of non-public networks enables examining security requirements and providing security by design. Therefore, further research and evaluation are required to develop the non-public network deployments compliant with the patient data privacy.

While e2e slicing is a key solution for non-public networks, a group authentication, security and exposure of the UEs within the enterprise need to be addressed. Particularly for security, the network slice ID advertisement should have an explicit broadcast zone otherwise increases possibility of privacy/security attacks. Furthermore, the dynamic aspect of 5G network slicing to automatically scale resources for NPN slices needs to be autonomously managed in the core, hence network slice instance lifecycle management load is required.

### C. Continuous Authentication

Whilst our review of related continuous authentication literature found applicable studies that could be used to authenticate the IoT devices that medical professionals use to access data, there are still some research challenges that exist within the field. One challenge is how to collect effective and reliable biometrics in the healthcare domain; in medical environments there may be a variety of different contexts that could affect the quality of the biometrics collected. Context variation has been shown to impede accuracy in some similar studies [17] so healthcare-specific contexts must also be considered and the detrimental effects mitigated (e.g., mobile keystroke dynamics in a ward compared to those from sitting in an ambulance).

## VI. CONCLUSIONS

Considering the technology advancements and security and privacy requirements in health sector due to operation sensitivity and personal data, communication service providers can play a key role advancing the services to improve a patient's outcome. The expanded security and compliance challenges that come with IoT, wearables and implantable sensors essential for continuous health monitoring demand improvement in existing security solutions by public networks. Considering different health applications and services, the presented non-public health enterprise network enables providing continuous customised services for patients and access to the data of these services for continuously authenticated medical staff. In this paper, we proposed the use of non-public network concept for healthcare service providers to improve the security and privacy of the patient data.

### ACKNOWLEDGEMENT

Partly funded by 5G Vinni EU H2020 research and innovation programme under grant agreement No 815279.